\documentclass[showpacs,aps,preprintnumbers,amsmath,amssymb]{revtex4}
\usepackage{graphicx,amssymb}
\begin{document}

\title{X-ray spectra of $3d$ metals of Heusler alloys and La$_{1-x}$Sm$_x$Mn$_2$Si$_2$ compounds}
\author{M.~V.~Yablonskikh, Yu.~M.~Yarmoshenko, E.G. Gerasimov,}
\author{V.S. Gaviko and E.Z. Kurmaev}
\affiliation{Institute of Metal Physics, Russian Academy of
Sciences-Ural Division, 620219 Yekaterinburg GSP-170, Russia}
\author{S.Bartkowski and M.~Neumann}
\affiliation{Department of Physics, Osnabr\"uck University,
D-49069 Osnabr\"uck, Germany}

\begin{abstract}
 The Mn x-ray emission spectra and x-ray photoemission spectra of
Mn-based Heusler alloys Co$_2$MnAl, Co$_2$MnSb and
La$_{1-x}$Sm$_x$Mn$_2$Si$_2$ compounds (x=0, 0.8) have been
measured and discussed in connection with a value local magnetic
moment at Mn site. The spectra peculiarities reflect also the
localization degree of $3d$ valence electrons of $3d$ metals in
considered compounds.
\end{abstract}


\maketitle

\section{Introduction}
 Magnetic properties of localized and
delocalized magnetics are usually bound with the peculiarities of
their electronic structure and found to be explained within the
framework of these terms.  Here x-ray emission (XE) and x-ray
photoemission (XP) spectroscopy is used to bring light on the
differences in $3d$ metal spectra of close to half-metallic
\cite{GrootHMFShreder7} Mn based Heusler alloys and rare-earth Mn
containing compounds both within general problem of strong
magnetism of the $3d$ metal alloys and compounds. Using the
samples well studied experimentally before by different techniques
we will discuss the difference at x-ray spectra of 3d metals in
connection with the value of the local magnetic moment and
different spin polarization of the valence electrons using the
x-ray spectroscopy. The selection of the samples will be explained
below. First let us pay attention to the techniques being used.

The x-ray spectroscopy is a well-known to be an element-selective
probe of electronic structure. That opens the principal ability to
provide information about the relation between a magnetic
properties and x-ray spectra, because it deals with a short-lived
states, so it is complementary with data obtained from static
magnetic measurements. The emission process accrues in a first
coordination sphere, therefore it is good to use that to
investigate the formation of local magnetic moment of $3d$ metal.
General techniques being used are the x-ray emission, x-ray
photoemission and x-ray absorption spectroscopy using the
circularly polarized energy-variable excitation source, that
allows to study a fine structure of electronic states and their
spin configuration. That could be combined with a tradition
techniques using non-polarized excitation with a fixed excitation
energy. Provided information gives a chance to distinguish
localized and itinerant magnetics containing $3d$ transition
metals by processing of x-ray spectra of $3d$ metals in different
types of magnetic alloys and compounds. Here we considered $3d$
transition metal spectra of Heusler Alloys (HA) (Co$_2$MnAl,
Co$_2$MnSb) and La$_{1-x}$Sm$_x$Mn$_2$Si$_2$ compounds (x=0, 0.8).


First attention to Heusler alloys  has been paid in 1903 when
Heusler \cite{Heusler1} discovered ferromagnetic behavior in
compounds of nonmagnetic elements. Such ternary alloys are
determined by the generic formulas X$_2$YZ or XYZ
\cite{webster1967,Parsons1995,FerromagneticBushow}
 For the Y-element
Heusler alloys usually contain Mn, the X element, for example, can
be X: Fe, Co, Ni, Cu, Pd, or Rh. For the Z element it is
considered Al, Ga, Si, Sn, Sb, or In. Crystallizing in the space
groups $L2_1$, Mn-based HA's have been found at most in a
ferromagnetic ground state. They offer the unique possibility to
study manganese compounds where the Mn atom has only other
transition X metals as nearest neighbors and non-transition Z
elements of group $III-V$ in the second coordination sphere.
Webster et al. and later on Robinson et al.
\cite{WebsterCryst,lmmr3} have investigated the crystal structure
of these alloys in detail. Furthermore, it has been shown that
some disorder is often appearing in Mn-based HA's and that in
total up to 10\% of the Mn atoms change places with elements from
a different sublattice. The neighborhood of Mn atoms is mainly
built by X and Z elements. Mn-based Heusler alloys have an
effective magnetic moment of about $\mu_{Mn}=3.0-4.0 \mu_B$
localized at the Mn site. Traditionally they are considered as
ideal systems with local magnetic moments.  First principle band
structure calculations \cite{Shreder6} for these alloys have shown
that in the electron system with spins directed along the
magnetization axis ($\uparrow$) the Mn-$3d$ states are occupied
and hybridized with the $3d$ states of the X atoms. In the
electron system with spins directed against the magnetization
($\downarrow$) local, not hybridized empty Mn-$d$ states have been
observed, located at $1.0-1.5$ eV above the Fermi level
\cite{Shreder6,Shreder5,IshidaBand,Williamsband83} Z atoms supply
\textit{sp} electrons, which take part in an indirect Mn-Mn
exchange and therefore determine the degree of occupation of
hybridized $pd$ orbitals. Further investigations on Mn-based HA's
~Refs. \cite{GrootHMFShreder7,HMFPtMnSbEZ3,Shreder66} predict the
existence of a half-metallic ferromagnetic (HMF) state, for
example, in NiMnSb, PtMnSb, PdMnSb, and PtMnSb. This means that an
energy gap at the Fermi level exists for one spin orientation and
a metallic behavior has been obtained for the other spin
direction. Other band structure calculations reveal that a HMF
ground state may also exist for Co$_2$MnZ alloys \cite{IshidaHFM}.
 Both neutron scattering measurements
\cite{HeuslerNeutron1,HeuslerNeutron2,HeuslerNeutron3} and
theoretical predictions \cite{Moria} show well defined local
moments $\mu_{Mn}$ (see Table.\ref{table1}). The local magnetic
moment at Co atoms is much smaller then at Mn (less then 0.5
$\mu_B$, \cite{WebsterCryst,lmmr1}. The remarkable feature of Mn
$3d$ valence band for alloys being considered is a theoretically
predicted pseudo gap for minority spin electrons, that makes
alloys to be close to class of half-metallic ferromagnets.

The RMn$_2$Si$_2$ compounds (R is rare-earth metal) crystallize in
the tetragonal ThCr$_2$Si$_2$ structure, where separate layers are
stacking along c-axis in the $-Mn-X-R-X-Mn-$ sequence. The
compounds are assumed to exhibit an unusual strong dependence of
Mn-Mn exchange interaction from the distance between Mn atoms in a
single layer \cite{CeLaMnSi,BrabersBuschow} . The relation of such
unusual Mn behaviour in RMn$_2$X$_2$ and peculiarities of their
electronic structure is suggested to be dependent of the
d$_{Mn-Mn}$ in the neartest neighborhood. In case of  d$_{Mn-Mn}$
<d$_c$ where d$_c$=2.85 - 2.87 $\AA$ is the critical distance for
the space localization of Mn $3d$ electrons, magnetic moments are
aligned primarily along the c-axis that leads to antiferromagnetic
order. In case of d$_Mn-Mn$ ferromagnetic order is appeared.
 The distance between
R(001) and Mn(100) planes is about $c/2=5.2\AA$, that exceeds the
one between R atoms ($d_{R-R}=a \approx 4.2\AA$) and Mn atoms
($d_{Mn-Mn}=\frac{a}{\sqrt{2}} \approx 2.8\AA$). Last value is
close to the critical $d_c \backsim $2.84 for space
localization-delocalization of Mn $3d$ electrons in the crystal
space according to predicted \cite{Goodenought}. The doping of the
second rare-earth metal in RMn$_2$Si$_2$ compoundes allows to
gradually change Mn-Mn distance from d$_{Mn-Mn}>d_c$ for
LaMn$_2$Si$_2$ to d$_{Mn-Mn}<d_c$ for SmMn$_2$Si$_2$
\cite{Evgen2002}

The system has an effective magnetic moment about
$\mu_{eff}=3.5-3.8 \mu_{B}$, that is similar to Heusler alloys,
although the value of $\mu_{Mn}$ is smaller and found to be in
range $2.3-2.5 \mu_{B}$. The value of local magnetic moment
$\mu_{Mn}$ changes from  0 to 1.45 $\mu_{B}$ with $x$ changing
from 1 to 0. The relation of such unusual Mn behavior in
RMn$_{2}$X$_{2}$ and peculiarities of their electronic structure
is suggested to be depended of the $d_{Mn-Mn}$ in the nearest
neighborhood.  In case of $d_{Mn-Mn}\approx d_{c}$, where
$d_{c}=2.85-2.87$ \AA ~is the critical distance for space
localization of Mn $3d$ electrons, magnetic moments align
primarily along the $c$-axis that leads to antiferromagnetic
order, and in case of $d_{Mn-Mn}>d_{c}$ the ferromagnetic ordering
appears. Note that in set RM$_{2}$X$_{2}$ doping by different $3d$
transition metals M=Fe,Co,Ni,Mn, only manganese contains the
non-zero local magnetic moment. Calculations \cite{LaMn2Si2Calcs}
show the non-strongly spin-polarized Mn DOS, unlike in Heusler
alloys. It was suggested that Mn $3d$ states of RMn$_2$X$_2$ have
been hybridized with X-$p$ states, in our case-with Si-$p$ states
due to close positions in a single layer. The direct Mn-Mn
exchange interaction prevails in a single layer of Mn atoms. The
one  between the nearest single layers  is realized by the La-$f$
and Sm-$f$ electrons hybridized with Mn $3d$ and also Si-$p$
electrons.

Finally we concentrates on the question of the influence of
localization degree of $3d$ electrons, the values of the local
magnetic moments of 3d metals at x-ray spectra. The Mn-Mn distance
in Heusler alloys exceeds the value for RMn$_2$Si$_2$ in factor
two. In other hand, the magnetic ordering is formed by indirect
intralayer exchange, that is opposite to direct exchange mechanism
in RMn$_2$Si$_2$. Calculations also predict the localization
difference in \textbf{k}-space \cite{LaMn2Si2Calcs}. These factors
defined the choice of selected materials. The principal
opportunity of using the spectroscopy to detect the influence of
an atom magnetic moment to X-ray spectra spectra had been shown by
van Acker et al.\cite{Acker}. The corresponding experimental data
have been obtained for a series of alloys and compounds containing
Fe. The energy splitting of the Fe-$3s$ band is essentially
different for each material, but without strong correlation
effects. From more recent experiments, in which polarized
radiation has been used it is evident that the dispersion as well
as the energetic position of the Fe-$3s$ band varies as a function
of the electron spin \cite{Fe3sMCDPhotoem}.


\section{Experimental details}
The specimens were prepared from a melt in an atmosphere of
purified argon, annealed at 720 $^\circ$C during 24 h  for Heusler
alloys and at the temperature $T=900 ^{\circ}C$ for
La$_{1-x}$Sm$_x$Mn$_2$Si$_2$ in argon and quenched in water.
According to X-ray diffraction patterns (Cu $K\alpha$) samples
were in single phase.

 X-ray photoelectron spectra (XPS) were measured using a PHI
5600ci spectrometer with an energy resolution of 0.35 eV with
monocromatized Al $K\alpha$ radiation. The compounds were crushed
\textit{in situ} in vacuum $1\cdot10^{-7}$ Torr. The surface
cleanlyness was tested by monitoring the C-$1s$ and O-$1s$ peaks.
During the measurements in vacuum $1\cdot10^{-9}$ Torr no
significant increase in the contamination was observed.

 X-ray emission spectra were obtained
using an x-ray spectrometer of the type RSM-500. The measurements
have been carried out at a voltage of U=4.5 kV, an anode current
of about 0.3-0.4 mA and $1\cdot10^{-7}$ Torr vacuum.  The energy
resolution in the XE spectra resulted in 0.7 eV. Last but not
least the rest contamination with oxygen has been estimated during
the heating procedure. This has been done by controlling the
intensity ratio of the lines O $K\alpha$ and Mn $L_l$ in the
corresponding XE spectra.

\section{Discussion}
\begin{figure}
\includegraphics[width=.7\textwidth]{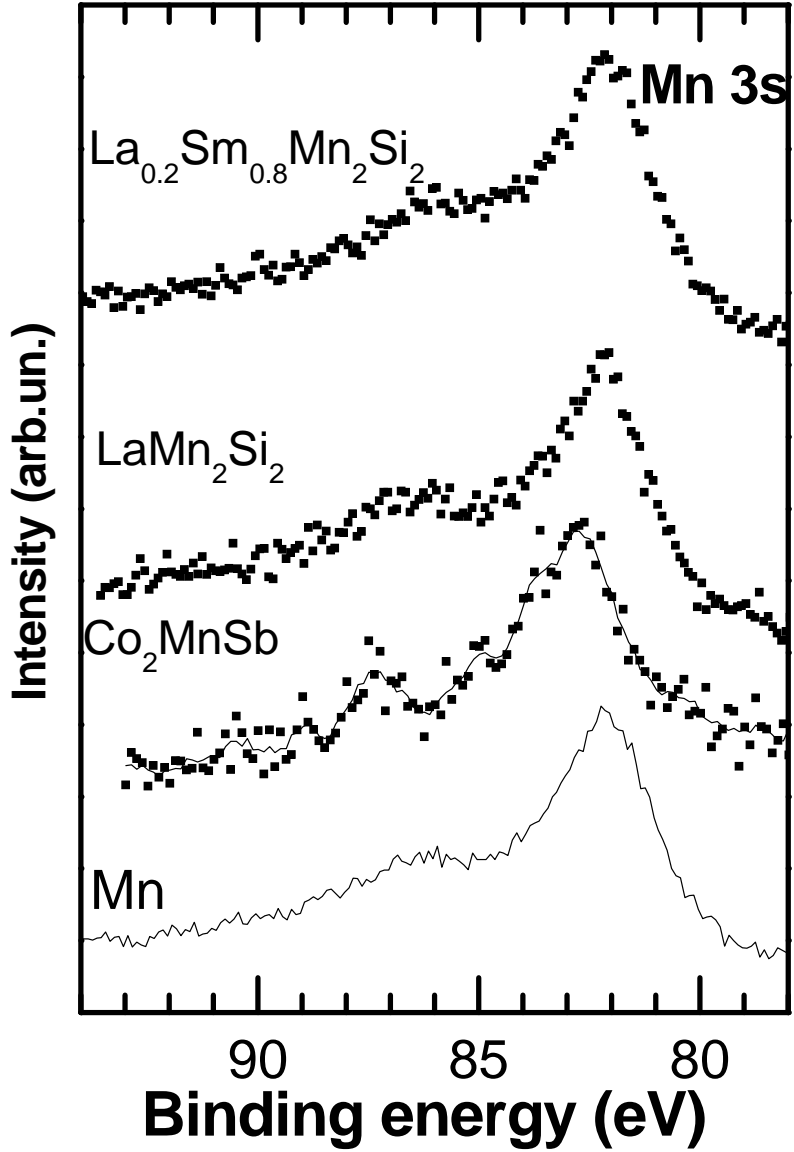}
  \caption{Mn $3s$ X-ray photoemission spectra.}
  \label{Mn3sfig}
\end{figure}

 The Mn 3s photoemission spectra of compounds and pure manganize are shown at
Fig.\ref{Mn3sfig}. The multiplet splitting is observed in form of
two peaks at the binding energies $E_B=83$ and $E_B=87.5$ eV for
pure Mn and RMn$_2$Si$_2$. No clear dependence of the spectra
shape and the local magnetic moment at Mn site is observed for
whole series of compounds, but roughly it is possible to
distinguish only small splitting of Mn $3s$ core level due to
interaction between $3d$ valence electrons and $3s$ core hole
created and the binding energy shift of Mn $3s$ XPS about 0.6 eV
toward high energy side.

\begin{figure}
\includegraphics[width=.7\textwidth]{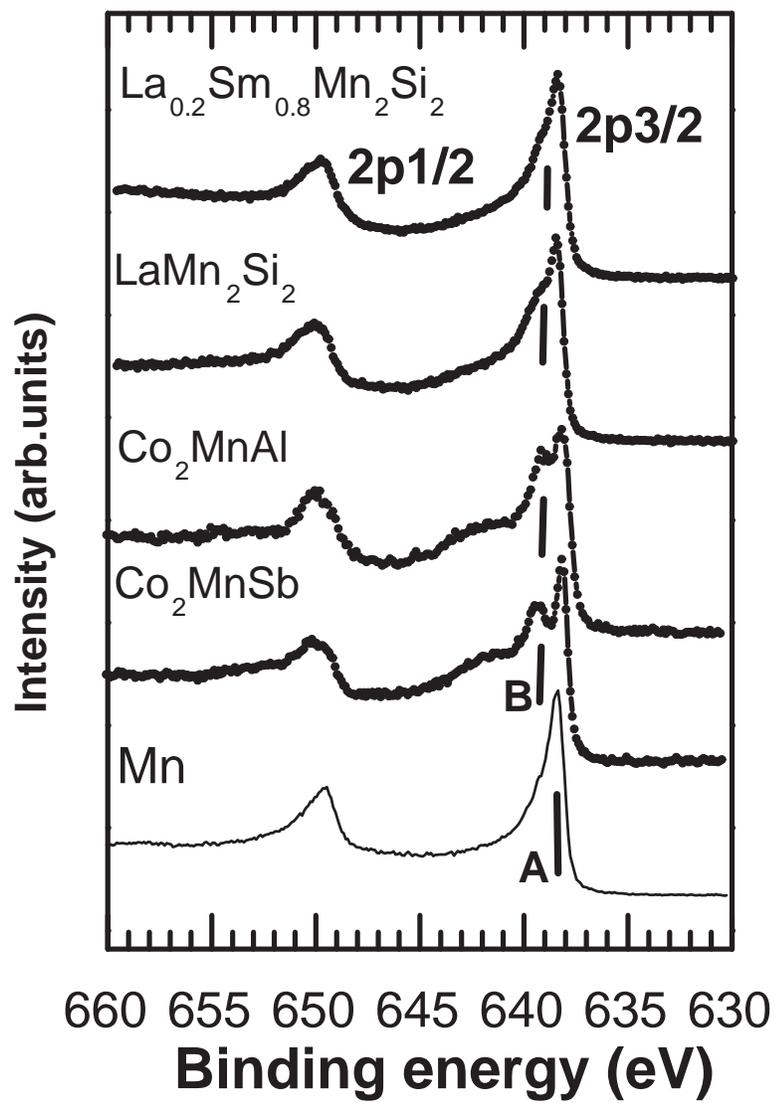}
  \caption{Mn $2p$ X-ray photoemission spectra.}\label{Mn2pfig}
\end{figure}

 Briefly, such effect is explained by the
difference in the energy  for Mn $3s^1_j$3d$^5S$ final state, that
comes to be result of the strong exchange interaction between Mn
$3s_j$ hole and strongly spin-polarized Mn $3d$ electrons in case
of Co$_2$MnSb. At Fig. \ref{Mn2pfig} the Mn $2p$ XPS have been
shown.

The spin-orbit splitting at the Mn $2p_{1/2}$ peak with the
binding energy $E_B=649.7$ eV and at the Mn $2p_{3/2}$ peak with
$E_B=638.5$ eV are observed clearly. Opposite to Mn $3s$
photoemission spectrum at Fig.\ref{Mn3sfig}, Mn $2p$ spectrum
differs from pure Mn for both series of compounds. The Mn
$2p_{3/2}$ peak splits at two, each position of is marked as $A$
and $B$. The energy position of peak $A$ is the same for all Mn 2p
spectra. The energy position of the peak $B$ is moving toward high
binding energies and the intensity is rising in line
La$_{02}$Sm$_{08}$Mn$_2$Si$_2$, LaMn$_2$Si$_2$, Co$_2$MnAl,
Co$_2$MnSb. For RMn$_2$Si$_2$ compounds the value of splitting is
0.7-0.8 eV and for Heusler alloys it is approximately 1 eV. No
energy shifts of Mn $2p$ XPS have been observed comparing to pure
metal spectrum. Peak of very weak intensity at the $E_B=641$ eV
appears at the edge of Mn $2p_{3/2}$ spectra due small amount of
oxygen at the surface.

\begin{figure}
\includegraphics[width=.7\textwidth]{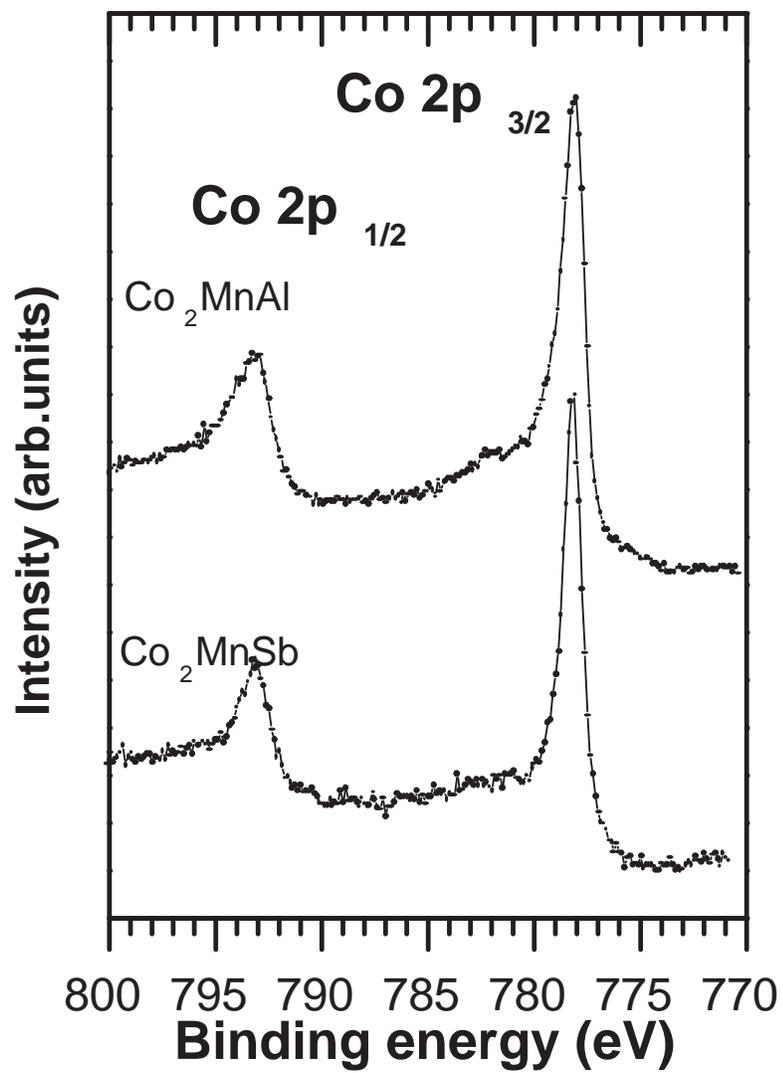}
  \caption{Co $2p$ X-ray photoemission spectra of Heusler alloys.}
  \label{Co2pfig}
\end{figure}

\begin{figure}
\includegraphics[width=.7\textwidth]{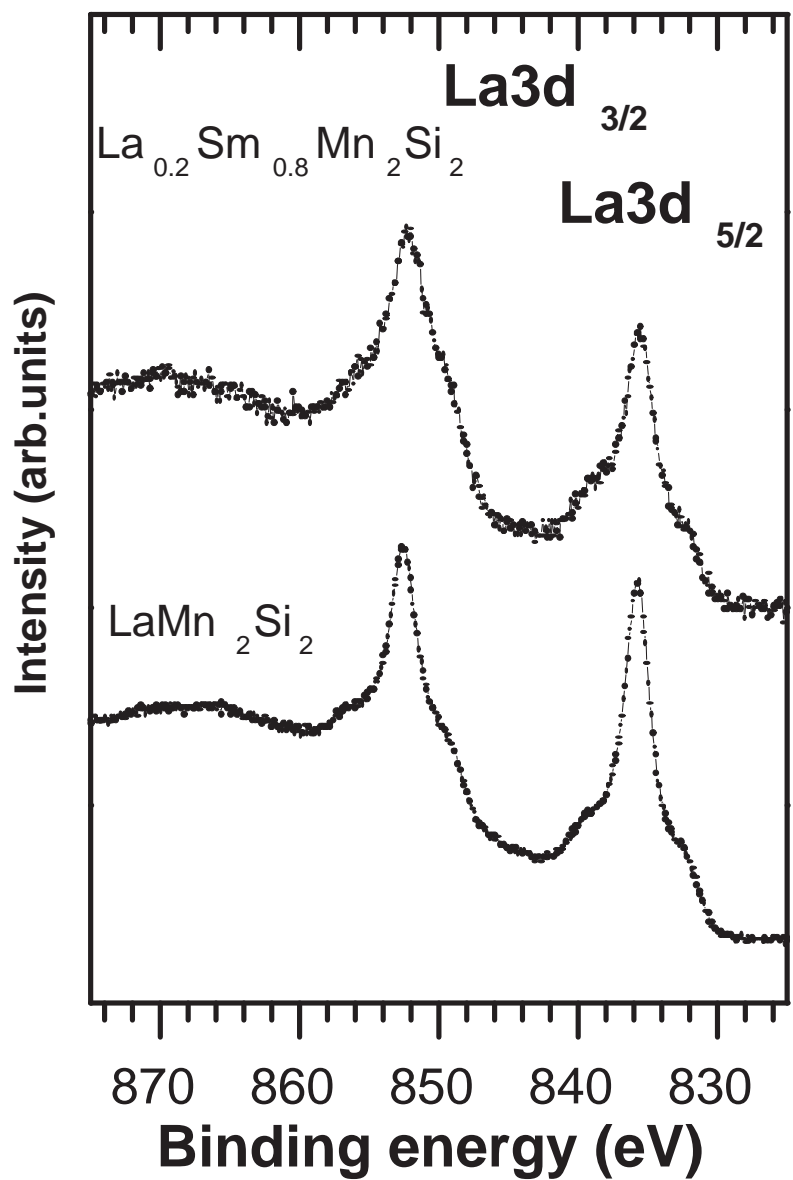}
  \caption{La $3d$  X-ray photoemission spectra of rare-earth compounds.}\label{La3den}
\end{figure}

 The Mn $2p_{3/2}$ energy splitting is connected to the strong
$2p-3d$ exchange interaction \cite{Yarmoshenko1} that correlates
with $\mu_{Mn}$. Such qualitative dependence is observed between
RMn$_2$Si$_2$ with $\mu_{Mn}=2.3-2.8$ and Heusler alloys with
$\mu_{Mn}$   see Table\ref{table1}. The absence of satellite peaks
and energy shifts at whole Mn $2p$ XPS allows to expect that
effects being observed for Mn $3s$ XPS are due to local magnetic
moment at Mn site, like in cases \cite{Acker,Fe3sMCDPhotoem}. That
also is in agreement with Co $2p$ and La $3d$ photoelectron
spectra at Fig.\ref{Co2pfig} and Fig.\ref{La3den} displaying no
significant changes.

\begin{figure}
\includegraphics[width=.7\textwidth]{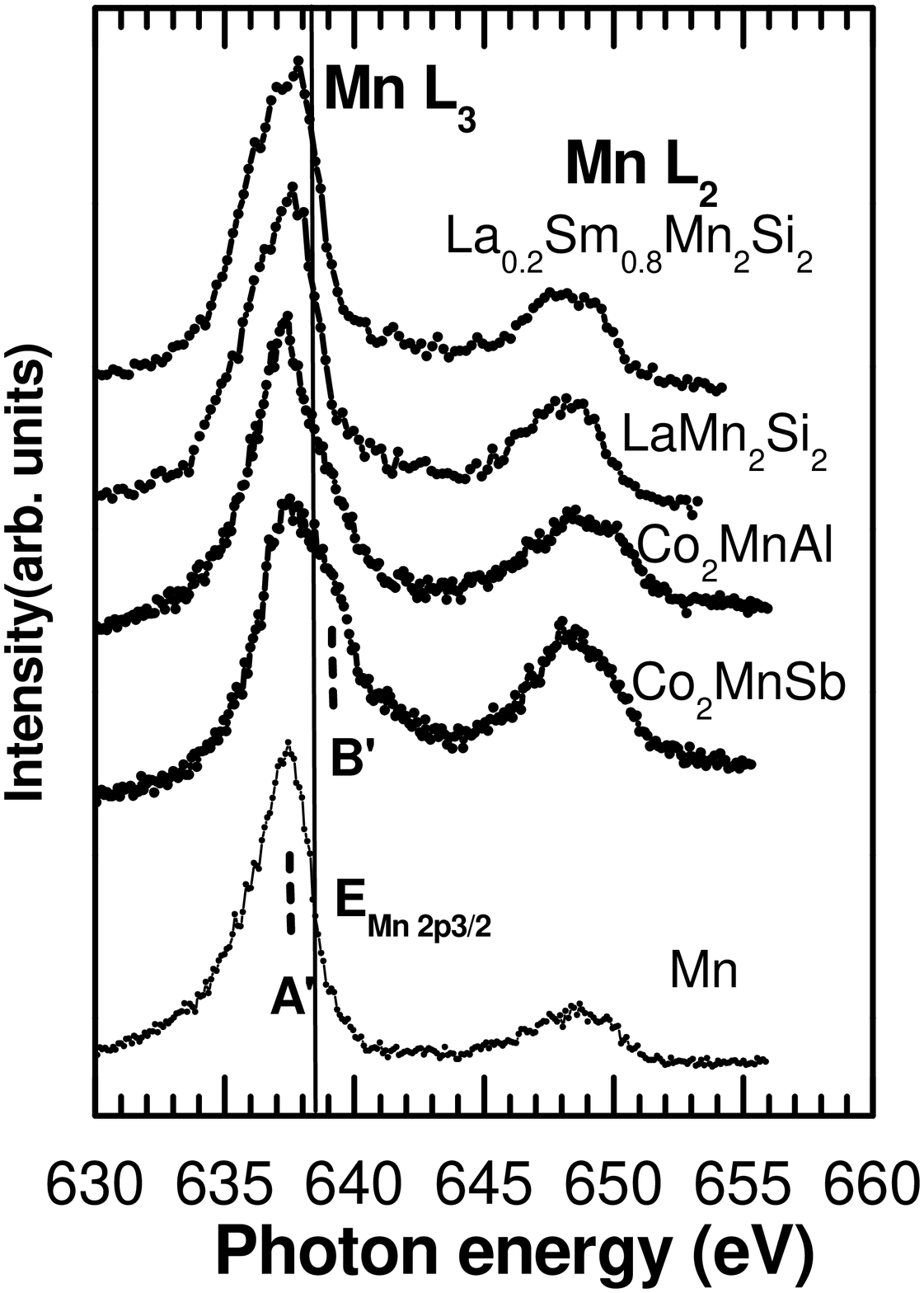}
  \caption{Mn $L_2,_3$  X-ray emission spectra of Heusler alloys.}\label{MnL23fig}
\end{figure}

  The Mn $L_2$,$L_3$ X-ray emission spectra corresponds to emission transition
of Mn $3d$ valence electron to $2p_{1/2}$ and $2p_{3/2}$ core hole
respectively. Comparing spectra between each other
(Fig.\ref{MnL23fig}) one finds two remarkable features. The height
of $I(L_2)/I(L_3)$ intensity ratio correlates with $\mu_{Mn}$. The
$L_3$ peak consist of two peaks $A'$ and $B'$, that is observed in
case of Heusler alloys and again is due to high localization of
strongly spin-polarized Mn $3d$ electrons \cite{OurMCDPReB}.

 In  x-ray emission spectra of $3d$ metals such a peak $B'$ located
toward high photon energies is explained by the influence of inner
Coster-Kronig transition \cite{HagueCoster,Duda3,Magnusson}, but
that is opposite to the high intensity ratio observed. Both the
presence of the peak $B$' with energy exceeding the Mn $2p_{3/2}$
core level binding energy and the $I(L_2)/I(L_3)$ high intensity
ratio are indicating the existence of pseudo gap for one of Mn
$3d$ spin-projection at valence band. That is characteristic for
the case of the strong localization of $3d$ electrons
\cite{PReB63-035106,PReB60-6428,OurMCDPReB}.

\begin{figure}
\includegraphics[width=.7\textwidth]{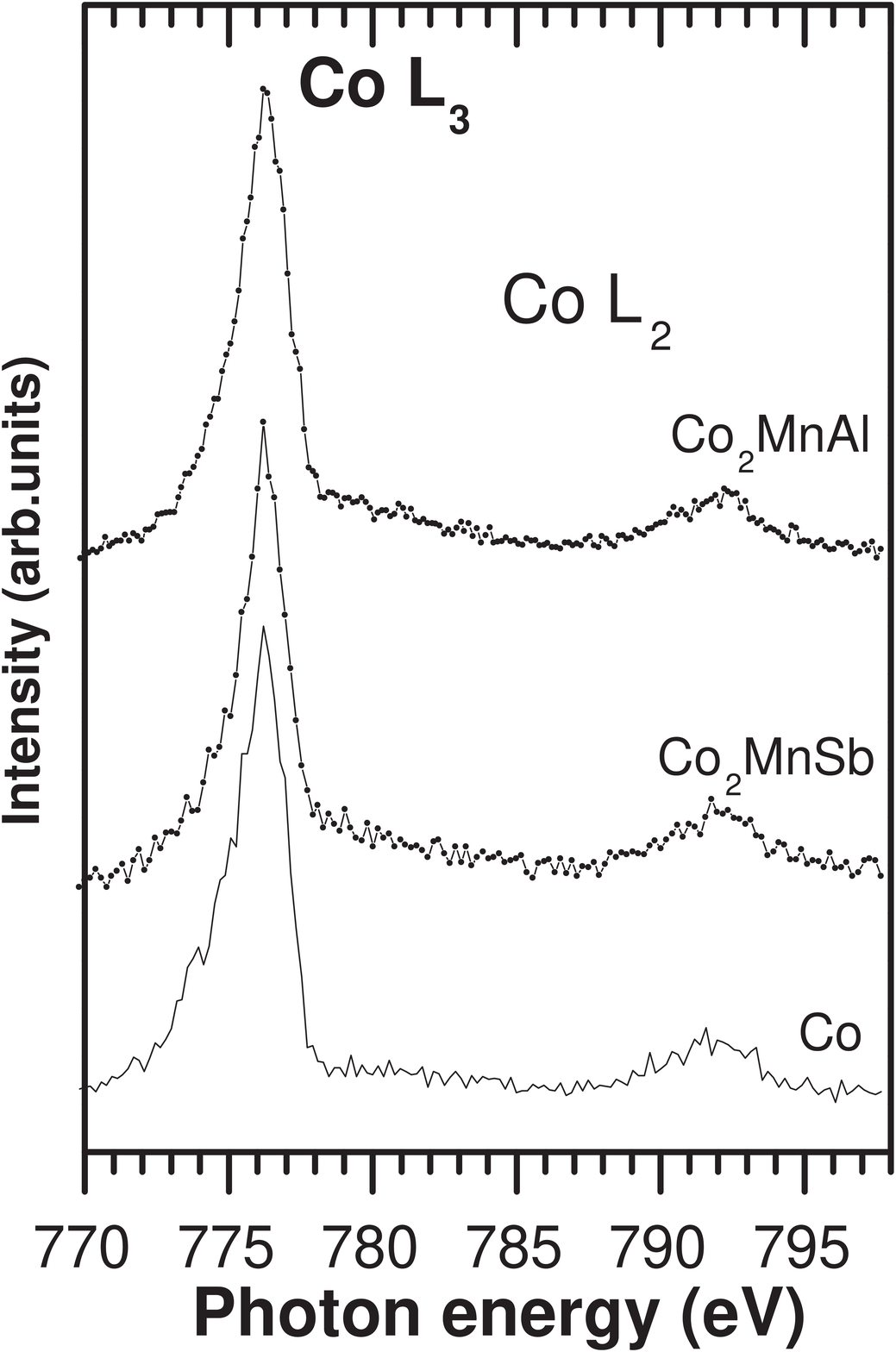}
  \caption{Co $L_2,_3$  X-ray emission spectra of Heusler alloys.}\label{CoL3en}
\end{figure}

 The Co XPS spectra of Heusler alloys and La, Sm spectra of core
levels are shown at Fig \ref{Co2pfig}, Fig.\ref{La3den}. There is
no difference between the pure metal spectra and one in compounds
. Co $L_2$,$L_3$ emission spectra resemble with one of pure Co,
that correlates with a small value (less then 0.5 $\mu_B$),
thought \cite{PReB60-6428}) of atom magnetic moment in compounds.
It is also confirmed by Co $L_2$,$L_3$ spectra exhibiting no
changes in comparison to the pure Co $L_2$,$L_3$ spectra (see
Fig.\ref{CoL3en}).

\section{Conclusion}
 The presented analysis gives an opportunity to discuss the degree
of valence electron localization for different compound with a
metallic type of conductivity. For example, analyzing the same of
spectra of metal oxides, x-ray spectra contains not only Coloumb
exchange interaction appears in x-ray spectra, but also
charge-transfer effects. Latter factor makes complicated the
interpretation of x-ray spectra. Both the degree of localization
of $3d$ valence electrons and type of chemical bonding is to be
reflected in x-ray spectra, that is observed at the example of Mn
spectra in compounds even using non-polarized and non-resonant
energy excitation source.


\bibliographystyle{prelsart-num}
\bibliography{All}

\newpage

\begin{table*}
\caption{Structural and magnetic properties of Mn-based compounds}
\begin{tabular}{|c|c|c|c|c|c|c|}
 \hline
material& structure  & $d_{Mn-Mn}$, nm & $\mu_{Mn}$,$\mu_{B}$ &$\mu_{eff}$,$\mu_{B}$ & $T_c$, K & $T_N$, K \\
 \hline
Mn & $\alpha-Mn$ &  & 1.9; 1.7; 0.6;0.2 &  &  & 95 \\
La$_{0.2}$Sm$_{0.8}$Mn$_2$Si$_2$   & ThCr$_2$Si$_2$ & 0.283 & 2.3 & 3.8 &  & 430 \\
LaMn$_2$Si$_2$  & ThCr$_2$Si$_2$ & 0.291 & 2.5 & 3.5 & 305 & 470 \\
Co$_2$MnAl  & $B2/L2_1$ & 0.407 & 3.01 & 4.01 & 693 &  \\
Co$_2$MnSb & $L2_1$ & 0.419 & 3.75 & 4.9 & 600 &  \\ \hline
 \hline
\end{tabular}
\label{table1}
\end{table*}

\newpage

\end{document}